# Single-Molecule Observation of Long Jumps in Polymer Adsorption

*Changqian Yu[a], Juan Guan[a], Kejia Chen[b], Sung Chul Bae[a], and Steve Granick[a,b,c,d],\**

[a]Departments of Materials Science and Engineering, [b]Chemical and Biomolecular Engineering, [c]Chemistry, and [d]Physics, University of Illinois, Urbana, IL 61801

*Corresponding author: sgranick@illinois.edu

**ABSTRACT**

Single-molecule fluorescence imaging of adsorption onto initially-bare surfaces shows that polymer chains need not localize immediately after arrival. In a system optimized to present limited adsorption sites (quartz surface to which polyethylene glycol (PEG) is exposed in aqueous solution at pH = 8.2) we find that some chains diffuse back into bulk solution and re-adsorb at some distance away, sometimes multiple times before either they localize at a stable position or else diffuse away into bulk solution. This mechanism of surface diffusion is considerably more rapid than the classical model in which adsorbed polymers crawl on surfaces while the entire molecule remains adsorbed. The trajectories with jumps follow a truncated Lévy distribution of step size with limiting slope -2.5, consistent with a well-defined, rapid surface diffusion coefficient over the times we observe.

The kinetics of polymer adsorption is fundamental to many polymer and biological applications. The classical picture depicts polymer chains to adsorb in three successive steps: (1) diffusion through the bulk to the surface, (2) surface attachment, and (3) chain relaxation into states that minimize conformational energy. The first step is rate-limited by the bulk diffusion coefficient while the third step is considered to reflect a process of chain flattening onto the surface, a type of "aging". However, commonly the second step of initial attachment to the surface is so rapid that there is no clear picture to describe it.[1-7]

In this field, the central dogma is that lateral mobility proceeds in the adsorbed state. Contrary to the traditional concept that adsorbed polymer crawls on surfaces while the entire molecule remains adsorbed,[1-4] the notion that interfacial molecules may diffuse by desorption-adsorption was first proposed by the theoretical model of O'Shaughnessy.[7] In this scenario, there are "strong adsorbers" such that re-adsorption is much faster than desorption, and other "weak adsorbers" that are released after some desorption time. As long-chain polymers display a broad distribution of bound fraction, some of the adsorbed chains will possess a low fraction of surface-adsorbed sites,[8-11] and these loosely-bound polymer chains should most easily overcome the sticking energy barrier and diffuse out into the bulk solution. Another related theoretical model comes from Monte Carlo simulations by Chakraborty[5] and Muthukumar.[6] In their vision, polymer chains tend to be detained at those points on the surface at which adsorption is strongest; there results a searching process of polymer-surface pattern recognition, the search for surface sites of strongest adsorption.

Here we test these ideas experimentally, for what is considered to be the first time from direct measurements at the single-molecule level. Experiments of this kind have become possible by single-molecule fluorescence imaging of individual polymer chains, combined with the accumulation of large datasets.

**RESULTS AND DISCUSSION**

The choice of experimental conditions was made as follows. Polyethylene glycol (PEG) adsorbs by hydrogen bonding to surface silanol groups. When the surface is highly protonated, there are many adsorption sites and multiple segments of a chain can absorb simultaneously and the chain will simply stick where they land. On the other hand, when the surface is highly deprotonated, the polymer will not adsorb. By tuning the pH, one can have a system where the surface is partially deprotonated, so that for certain polymer chains only a fraction of the segments on it will adsorb and the chain can desorb (Figure S1). To avoid the complication that some previously-adsorbed chains might block the arrival of new chains, we studied adsorption onto an initially-bare surface. Data in this paper refer to PEG ($M_w$ = 40 kg-mol$^{-1}$), unless noted otherwise.

Our exploratory experiments showed essentially zero surface mobility of this PEG polymer at pH < 7.0, when there is copious hydrogen bonding of PEG to surface silanol groups, and virtually no adsorption at pH > 8.8. This is consistent with previous ensemble averaged adsorption measurement through optical reflectivity.[12] We avoided experiments at the highest limit of workable pH, as under this condition the surface mobility was too rapid to be imaged with sufficient resolution. The optimized working region was found to be pH = 8.2 (± 0.1), as sketched schematically in Figure S1, at which it is estimated that 10 - 20% of surface silanol groups are deprotonated.[13] Note that the density of silanol groups on silica is ~5 per nm$^2$.[14] Figure S1 shows a hypothetical adsorbed chain with "loop-tail-train" conformations, the "train" segments adsorb to those random locations on the surface with protonated silanol groups, while the intervening segments are spanned by loops.

In this situation, the overall mobility of adsorbed PEG was very slow, as we measured in exploratory experiments using fluorescence recovery after photobleaching (FRAP). The deduced diffusion coefficient of adsorbed chains was ~0.8 × 10$^{-3}$ μm$^2$-s$^{-1}$ (Figure S2), with 10 nM polymer in the solution, a condition under which the fractional surface coverage is expected to have been 80 - 90% as this concentration of PEG is high enough to saturate protonated silanol surface sites.[12, 15] This diffusion coefficient is 5 orders of magnitude slower than the bulk diffusion coefficient of ~35 μm$^2$-s$^{-1}$ measured by fluorescence correlation spectroscopy (FCS), shown in Figure S3. However, a FRAP measurement gives only the ensemble averaged value. Moreover FRAP is not feasible to measure at very low surface coverage to probe how single polymer chains interact with the surface.

Single-molecule imaging shows that the underlying data are bimodal: the vast majority of polymer chains (~90%) were immobile on the experimental time scale of ~100 s, while a minority translated from point to point by jumps. Figure S4 presents the step size distribution of the former evaluated over 0.1 s, which was well fitted by a two-dimensional Gaussian distribution with the experimental spatial resolution of ~25 nm as the width at half maximum (inset of Figure S4 and Movie S1). However, the minority ~10% population shows trajectories containing occasional long jumps, as illustrated in Figure 1(a) and Movies S2 through S5. Some trajectories present multiple jumps.

The waiting time before a jump was broadly distributed, both between different jumps and between different polymer chains. Figure 1(a) illustrates some of the patterns that we observed. In trajectory 1, the chain displayed four long jumps before finally became immobile. Trajectories 2 and 3 illustrate chains that were immobile for tens of seconds at multiple spots. Trajectory 4 depicts a chain that executed numerous consecutive steps, and 5 experimental images at successive times in this trajectory are illustrated in Figure 1(b).

For further quantification, the mean-square displacement of the same four trajectories is plotted against time on log-log scales in Figure 1(c). All display, at long times, the slope of unity characteristic of Fickian diffusion, but at short times trajectories 2 and 3 are sub-diffusive, reflecting the long pauses already evident in the raw data. Trajectories 1 and 4 are also subdiffusive, but in those trajectories the pauses were shorter, and therefore the overall mobility was dominated by long jumps. Plotting displacement against time in Figure 1(d), we observe stick-slip motion: intervals of constant position separated by long-distance jumps.

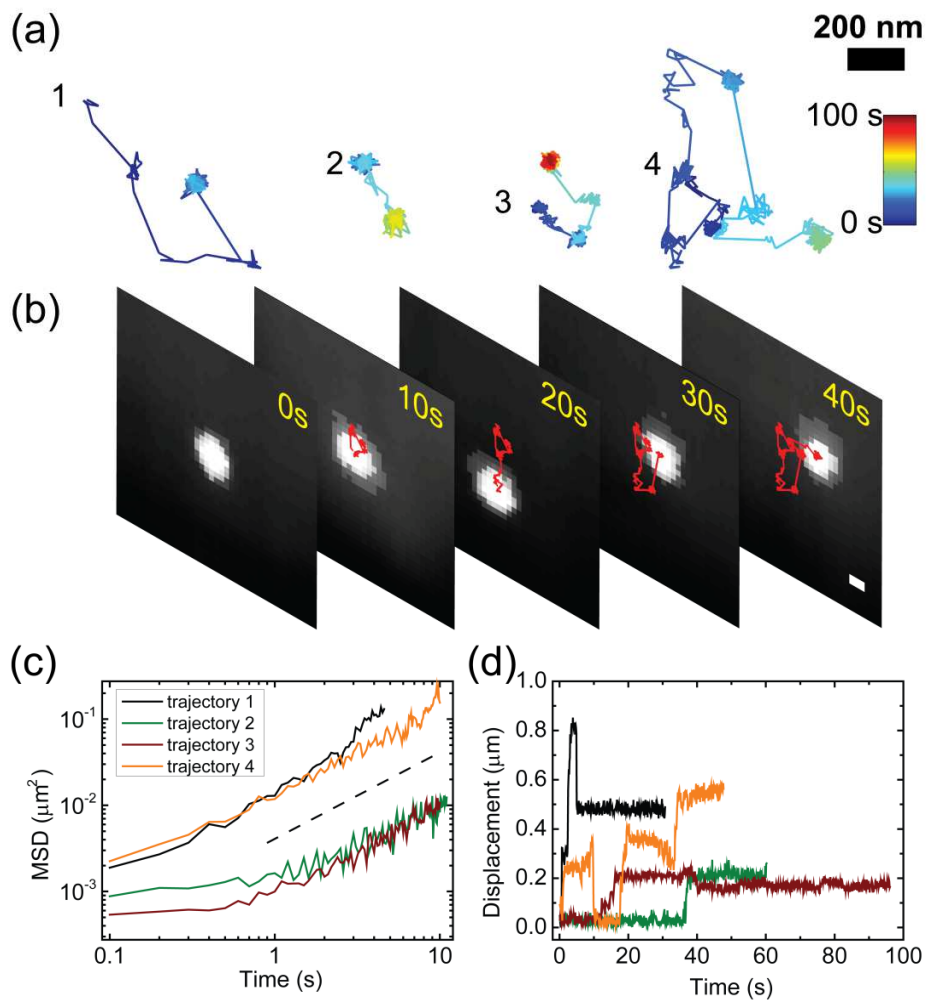

**Figure 1.** Analysis of individual trajectories. (a) Four representative trajectories with color-coded time, blue to red up to 100 s, color bar on the right. The time step is 0.1 s. (b) Five images at successive times during a trajectory (red line) at the indicated times up to 40 s. The diffraction-limited bright spot (after denoising) of a labeled dye molecule represents one polymer chain. The scale bar is 200 nm. (c) The mean-square displacement (MSD) is plotted against time on log-log scales. The dashed line has slope of unity. (d) Linear displacement plotted against time. Plateaus on each trace represent long trapped states.

These data suggest the hypothetical scheme of polymer adsorption summarized in Figure 2. Depending on where the chain lands, its mobility can be quenched immediately by forming many segmental bonds to the surface (a1), or remain mobile (a2) if only weakly adsorbed. The conformation entropy of chain fluctuations presumably encourages detachment. After chains detach, they may re-adsorb (b1 and b2) or diffuse back into the bulk solution (c). The theoretical models are consistent with this view qualitatively,[5-7] although they were not specifically concerned with lateral jumps or their distributions. However, surface diffusion whose rate is dictated by long jumps is known in systems other than polymers. For example, time-resolved STM has shown long jumps of metal atoms and also of organic molecules on metal surfaces.[16, 17] At a solid-liquid interface, fluorescence-based measurements of surfactant adsorption tell a similar story of jumps,[18] though the jump length was interpreted to follow a Gaussian distribution. The behavior we observed here is unequivocally not Gaussian, however, as we now describe.

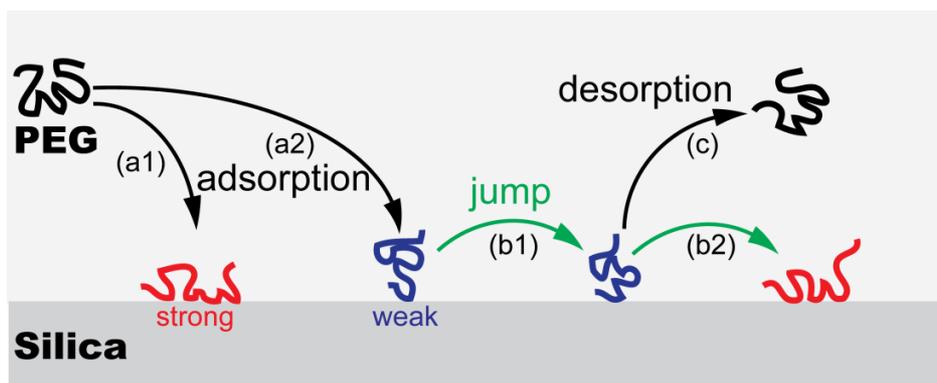

**Figure 2.** Hypothetical scheme of polymer jumping by an adsorption-desorption mechanism. (a1) Immediate adsorption without further searching; (a2) Weak, transient adsorption; (b1) Jumping from one surface spot to another by desorption/readsorption; (b2) Repeated jump to a third surface spot; (c) Alternatively, desorption into the bulk solution.

By accumulating thousands of similar trajectories, statistical quantification of the desorption-adsorption can be resolved, despite their rare occurrence in single trajectories. In Figure 3(a), for 100 representative trajectories the displacement over 0.1 s (frame-to-frame) is plotted against elapsed cascade time. The intermittent jumps appear as bursts in a background of pauses. More fundamentally, we rationalize that a desorbed chain at the vicinity of surface should have equal probability to diffuse back onto the surface or into the bulk solution, which means the probability of N jump events occurring within one trajectory will scale as $(1/2)^N$. Considering a step larger than 0.12 µm as a jump event as pauses are locally confined to a region size of ~0.05 µm (Figure S4) due to uncertainty of measurement, we find that the number of such jumps goes consistently as scaled probability $\sim(1/2)^N$ (Figure 3(b)). During jumps, chains are transiently desorbed with equal probability to diffuse into bulk solution or return to the surface.

The probability distribution of step size during 0.1 s is plotted in Figures 3(c) and (d) on semi-logarithmic and double logarithmic scales, respectively. This power-law distribution differs fundamentally from the Gaussian and exponential distributions noted in earlier work from this laboratory for other systems;[19] a power-law distribution is inherently more heterogeneous. The phenomenological power law is $\lambda = 2.5$

$$G(r, t=0.1s) \sim r^{-\lambda} \qquad (1)$$

Note that displacements below 100 nm (we attribute them to localized diffusion in the adsorbed state) deviate from this power law. We also include in Figure 3(d) consistent data for another molecular weight of polymer, $M_w = 10.8$ kg-mol$^{-1}$. The same power law distribution is robustly reproduced and the lower molecular weight contributes a slight right-shift to larger displacement due to the faster mobility in the bulk.

Such power-law diffusion is known in other systems. For example, molecular dynamics simulations revealed that adsorbed gold nanoclusters on graphite follow a flight-length power-law distribution with exponent -2.3,[20] close to what we observe here. Random walks of equal elementary step size encountering a planar boundary give a Cauchy distribution which asymptotically depicts flight length with similar power law.[7, 21, 22] More generally, Mandelbrot[23] linked random walks with large jumps to Lévy flights in fractal systems, with power-law in the range $1 < \lambda < 3$.[24] Mathematically it is known that when $\lambda \geq 3$, the distribution reverts to Gaussian in the limit of large numbers based on the central limit theorem,[24] so the process is consistent with Fickian diffusion. Power-law Lévy distributions are also known in numerous other systems, including marine predator foraging,[25] human mobility,[26] and the transport of DNA-binding proteins along DNA chains, which speeds up DNA transcription.[27] While the present system is not immediately related to searching strategy discussed in those biological contexts, the power-law distributed jump sizes in this polymer system may function as an effective search mechanism for strong adsorption sites. We presume that the search is for stable surface attachment, as a case for polymer-surface pattern recognition,[5, 6] but no quantitative explanation of the mechanism is offered at this time.

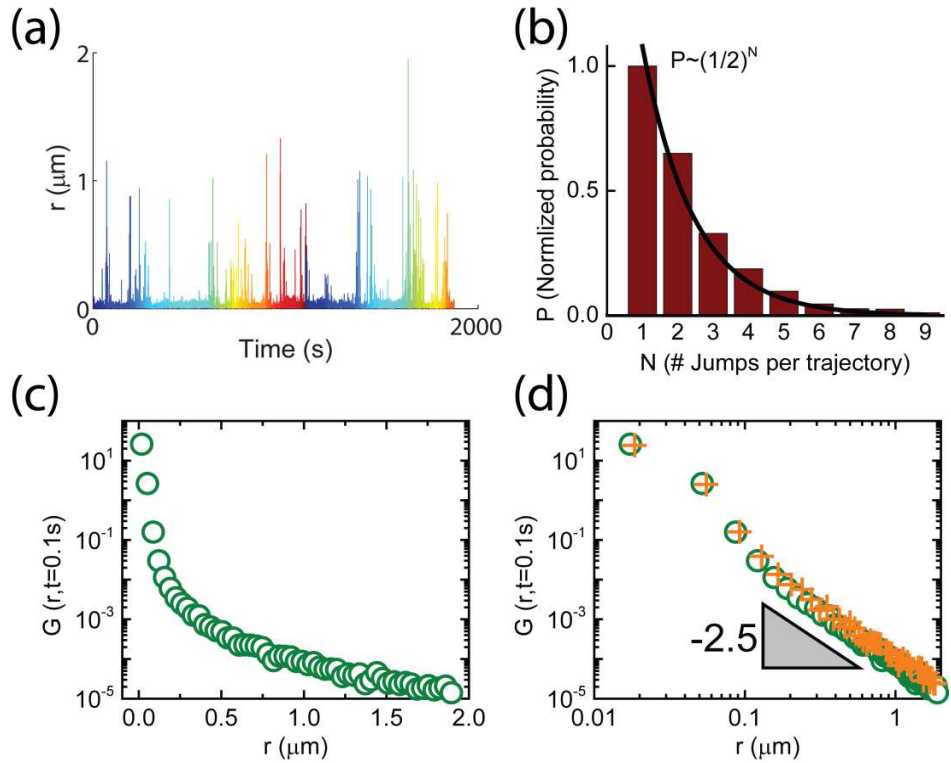

**Figure 3.** Statistical analysis of jump events. (a) Jumps observed in frame-to-frame (0.1 s) displacement from 100 representative trajectories with different trajectories coded with different colors. Time is cascaded by different trajectories. (b) Relative probability (P) of number of jumps (N) in a trajectory up to length of 100 s (jumps r > 0.12 µm), consistent with the solid line of $(1/2)^N$. (c) Jump size distribution inferred from thousands of trajectories; relative abundance is plotted logarithmically against linear jump size. (d) Same as (c) but on log-log scales. The orange cross marks show data for PEG of another molecular weight, $M_w = 10.8$ kg-mol$^{-1}$. The limiting slope is -2.5.

To find the connection between polymer diffusivity and power-law step size distribution with large jump events, we analyzed all polymer chain trajectories. In Figure 4(a), the mean-square displacement is plotted against time on log-log scales, having separated trajectories into two populations with or without displaying jumps. Jump-free trajectories have nearly zero displacement. During the interval up to 1.0 s, the mean-square displacement of trajectories with jump motion is linear in time, shown as Fickian diffusion with apparent diffusion coefficient ~2.5 × $10^{-2}$ $\mu m^2$-$s^{-1}$. Note that as these trajectories contain no jumps longer than 2 µm, this truncated power-law distribution has finite first and second moments.[28, 29] This apparent diffusion coefficient is ~30 times more rapid than the ensemble-average surface diffusion measured from FRAP, and ~1000 times slower than for these chains in bulk solution.

Further analysis follows by considering the model of continuous time random walk (CTRW),[30] according to which the distribution of displacement follows $G(r) \sim r^{-1-\beta}$ and the distribution of the residence time before jumps is $P(\tau) \sim \tau^{-1-\alpha}$. Fitting to these relations, it follows that $\beta \approx 1.5$ is implied by the step size distribution, and $\alpha \approx 0.8$ by the residence time distribution in Figure S5. The expected identity $2\alpha \approx \beta$ satisfies the linear mean-square displacement that we observe. We also notice that probability distributions of step size collapsed onto a master curve after normalizing r by $\sqrt{t}$ (Figure 4(b)), also indicating MSD ~ t during the time interval up to 1 s. It is possible that tracked immobile trajectories belong to 'crawling-like' motion along the surface. Using FRAP, we measured the ensemble-averaged diffusion coefficients for adsorbed PEG, with and without the presence of polymer in bulk solution. From the comparison (Figure S2) we see at least 10 times slower diffusion for the latter case, when bulk polymers had been rinsed out of the system. This result further supports the proposition that desorption/adsorption events speed up interfacial polymer diffusivity.

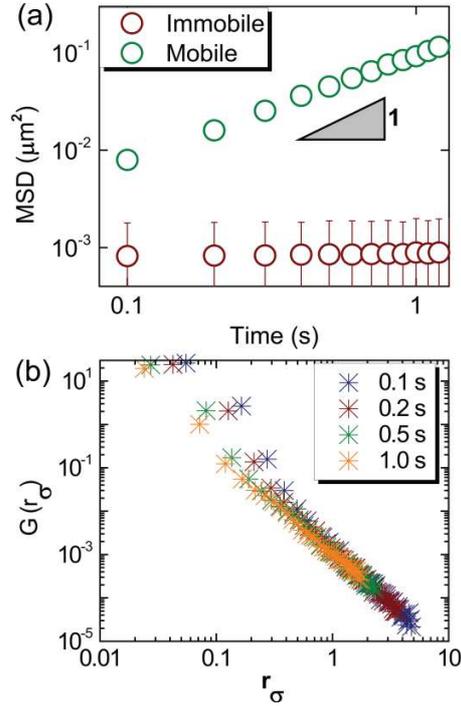

**Figure 4.** Time evolution of power-law tail in displacement distribution. (a) The overall mean-square displacement (MSD) of trajectories with and without jumps, labeled "mobile" and "immobile" in the figure. Within the tracking resolution, the latter has zero mobility. (b) Master curve of the probability distribution after normalizing displacement by the square root of time step, $r_\sigma = r/\sqrt{t}$. The data collapse with delay time 0.1 s, 0.2 s, 0.5 s, and 1.0 s, with the same limiting slope of -2.5 on log-log scales.

## CONCLUSIONS

The present experiment raises fundamental questions concerning how a single polymer chain adsorbs onto a solid surface from dilute solution. Polymer chains, resolved by single-molecule fluorescence imaging, can execute multiple jumps when they arrive at the solid-liquid interface, rather than immediately localize to discrete surface locales. Contrary to the traditional concept that adsorbed polymers crawl on surfaces while the entire molecule remains adsorbed,[1-4] these loosely bound chains can diffuse rapidly mediated by desorption followed by re-adsorption. The intermittent jump events we report statistically follow a truncated Lévy distribution of step size with limiting slope -2.5, contributing to exceptionally rapid surface diffusion of those chains over the times we observe.

## METHODS

End-reactive polymer, amino-terminated methoxy polyethylene glycol (mPEG-NH$_2$, M$_w$ = 40 kg-mol$^{-1}$, M$_w$/M$_n$ < 1.08, Nanocs) was allowed to react with ATTO488 (ATTO TEC) dye, using standard synthetic procedures. This organic dye was selected, after screening numerous other dyes, because of its exceptional brightness and photo-stability, which allowed us to measure single-molecule trajectories lasting up to tens of seconds.[31] The influence of chain length on interfacial polymer dynamics was investigated by PEG with a second molecular weight (mPEG-NH$_2$, M$_w$ = 10.8 kg-mol$^{-1}$, M$_w$/M$_n$ = 1.08, Polymer Source).

Unreacted dye was removed by passing the solution through a commercial dye removal column (Thermo Scientific) at least 5 times. Fluorescence correlation spectroscopy (FCS) measurements gave a clean single diffusion coefficient ~35 µm$^2$-s$^{-1}$ in aqueous solution (Figure S3), confirming the absence of residual non-reacted dyes. Solutions were prepared at a concentration of several tens of pM with 10 mM Na$_2$HPO$_4$/NaH$_2$PO$_4$ to buffer the solution at pH = 8.2 ± 0.1. To scavenge oxygen radicals, 5 mM sodium ascorbic acid was added. The adsorbing surface consisting of quartz coverslips (SPI Supplies) was hydroxylated using a cautious application of Piranha solution followed by treatment in an oxygen plasma cleaner (Harrick Scientific). It was our experience that this protocol resulted in the total removal of fluorescent contaminants; by direct imaging, their successful removal was verified before each experiment. In control experiments, the free dye, negative in charge, did not adsorb to the (negatively charged) quartz surface, thus further assuring that our fluorescence measurements truly represented polymer chains, not free dye.

An EMCCD camera (Andor iXon) was used to acquire the single-molecule data presented above. Images were acquired at 10 frames per second through an oil immersion objective (100×, NA = 1.45, Zeiss) using total internal reflection fluorescence (TIRF) microscopy. Signal-to-noise ratio of video images was significantly enhanced by image denoising algorithms[32] and trajectories were linked between frames by MATLAB code written in-house (see Movies S2 and S5).[33] The denoised images offer better spatial resolution without altering the physical interpretation of polymer dynamics.[34] Based on data presented in Figure S4, local changes in 2D positions of labeled polymer were resolved with 25 nm precision. The threshold to link points between consequent frames is set at 2 µm, as the average displacement during 0.1 s in bulk solution is expected to be $(4 \times 35 \ \mu m^2/s \times 0.1 \ s)^{1/2} \approx 3.7$ µm. The frame interval (0.1 s) was optimized to capture an entire jump event while short enough to measure surface residence time. Jump events were found to be so transient that we never observed successive jumps in two consecutive frames. All trajectories were overlaid with the original movie files and confirmed by visual inspection. Our methods of fluorescence recovery after photobleaching (FRAP) measurement have been described elsewhere.[35]

**Conflict of Interest:** The authors declare no competing financial interest.

**Acknowledgements:** This study was supported by the taxpayers of the U.S. through the National Science Foundation (Polymers Program), DMR-0907018.

**Supporting Information Available**: Experiment Scheme, FCS curve of PEG bulk diffusion, FRAP measurement of the ensemble averaged diffusion coefficient of adsorbed polymer, the probability distribution of step size for immobile surface-bound polymers, distribution of surface residence times, the movie for position tracking of immobile polymers, movies for four example trajectories with intermittent long jumps. This material is available free of charge *via* the Internet at http://pubs.acs.org.

Table of contents (TOC) graphic

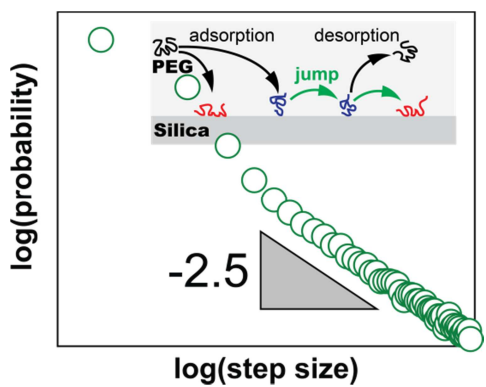

# Supporting Information

# Single-Molecule Observation of Long Jumps in Polymer Adsorption

*Changqian Yu[a], Juan Guan[a], Kejia Chen[b], Sung Chul Bae[a], and Steve Granick[a,b,c,d],*

[a]Departments of Materials Science and Engineering, [b]Chemical and Biomolecular Engineering, [c]Chemistry, and [d]Physics, University of Illinois, Urbana, IL 61801

*Corresponding author: sgranick@illinois.edu

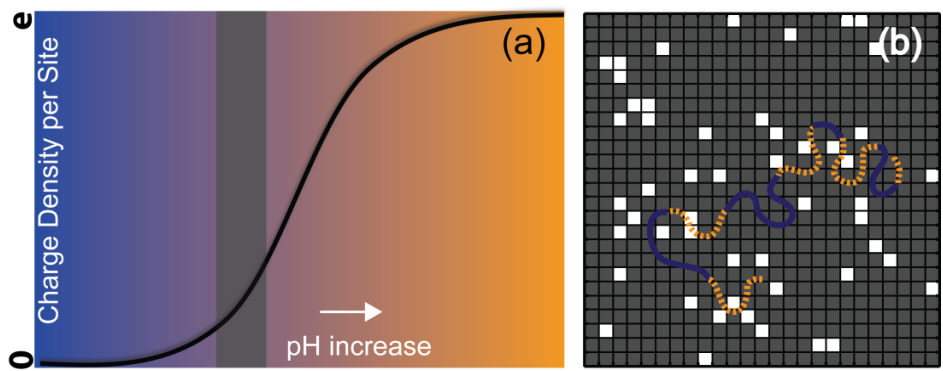

**Figure S1.** Experimental scheme. (a) Surface charge density schematically plotted against pH for quartz immersed in aqueous solution. Silanol groups on quartz deprotonate as pH increases and so charge increases. Color gradient (from blue to orange) represents relative polymer mobility: immobile at low pH while unlikely to adsorb at very high pH. The gray region denotes pH = 8.2 (± 0.1), a condition with 10 - 20% silanol groups deprotonated, at which experiments in this paper were performed. (b) Schematic representation of a quartz surface (25 nm × 25 nm) with 10% deprotonated silanol groups (shown as white spots; for simplicity, one spot represents 5 sites per nm$^2$) to which polymer segments cannot adsorb. Surface-hugging "trains" of polymer segments (blue) bind to the surface while "loops" and "tails" (orange) span intervening surface sites to which polymer segments do not stick.

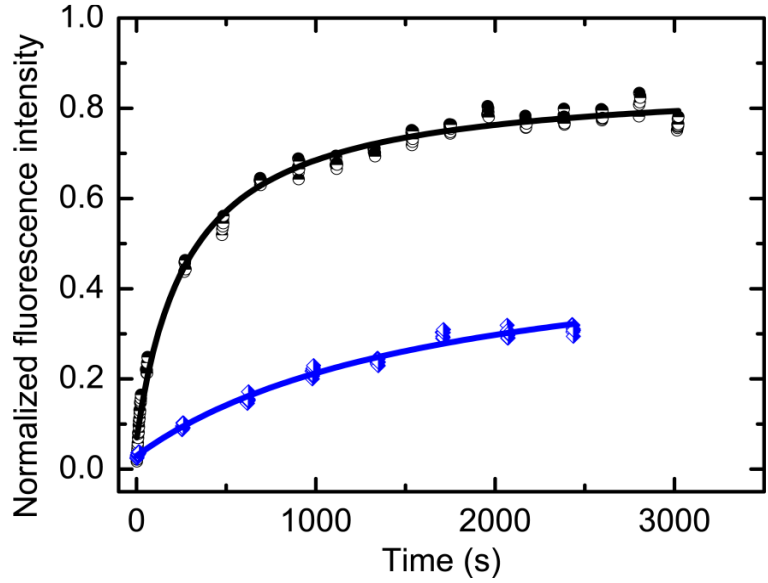

**Figure S2.** A FRAP measurement of the ensemble-averaged diffusion coefficient of adsorbed polymer at fractional surface coverage of 80 - 90%. Recovery is considerably faster with polymer in free solution (concentration 10nM, black circles) than if polymer is previously rinsed out of free solution (blue diamonds). Solid lines are fitted by the equation $I(t) = I_0 \times [1 - \frac{A}{1 + \frac{t}{\tau}}]$, where $I_0$ and $A$ are fitting parameters which represent the final recovered intensity plateau and the extent of photobleaching in the bleached spot.[1] The diffusion coefficient, $D$ can be obtained by $D = \frac{\omega^2}{4\tau}$, where $\omega \approx 1 \mu m$, is the diameter of the diffraction limited bleaching spot. Before and after rinsing, the characteristic diffusion time $\tau$ is ~300 and 3000 s, respectively.

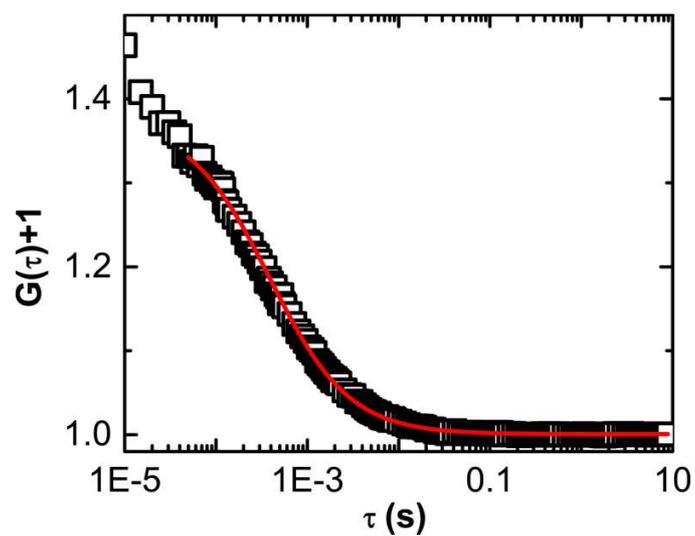

**Figure S3.** The diffusion coefficient (D) of 40K PEG in bulk aqueous solution at pH = 8.2, measured using one-photon fluorescence correlation spectroscopy (FCS). The autocorrelation function is plotted against logarithmic time lag and fit to the model of a single diffusing species (red), implying D ≈ 35 $\mu m^2 \cdot s^{-1}$.

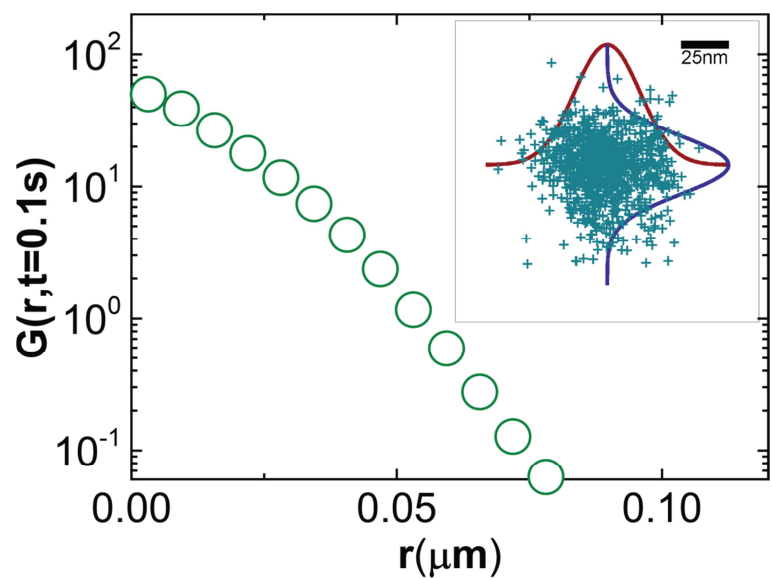

**Figure S4.** The probability distribution of step size, evaluated over 0.1 s, for immobile surface-bound polymers, plotted on semi-log scales. These data include thousands of trajectories with one example shown in the inset. Tracking resolution (inset) is inferred from Gaussian fitting to be 25 nm (full width at half maximum).

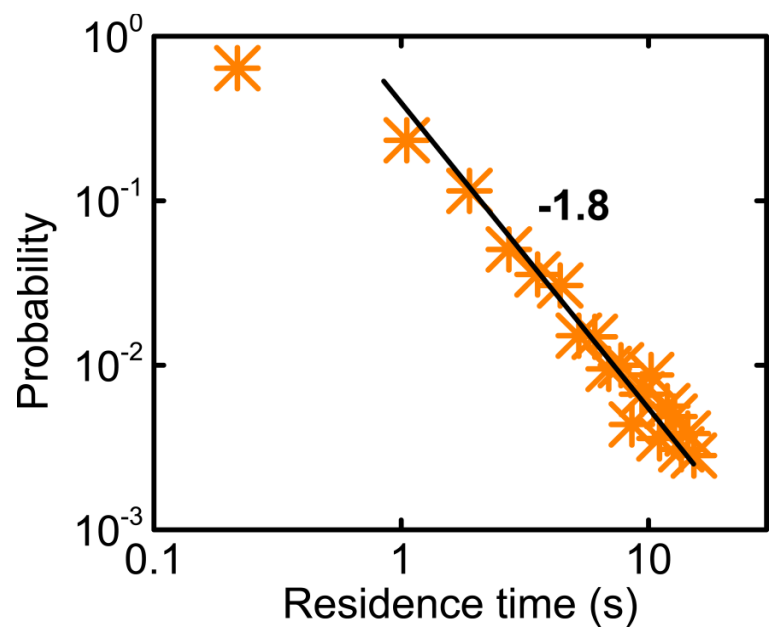

**Figure S5.** Distribution of surface residence times when considering trajectories with displacement less than 0.12 μm.

## Movie legends

**Movie S1.** Position tracking (after denoising) of an immobile polymer molecule. Blue star denotes 2-D Gaussian fitted position in each frame. Scale bar is 100 nm and the movie is played at 3× real time.

**Movies S2 through S5.** Four trajectories, shown in Figure 1(a), steps linked by red lines, of polymers, displaying intermittent long jumps. Movie S2 and S5 display in addition side-by-side comparison of the raw data and images after denoising. The denoised images offer better tracking resolution of polymer position without distorting dynamic information. Scale bars are 200 nm.